\documentclass[utf8]{FrontiersinHarvard}

\usepackage{url,hyperref,lineno,microtype}
\usepackage[onehalfspacing]{setspace}

\usepackage{graphicx}
\usepackage{booktabs}
\usepackage{amsmath,amsfonts,amsthm,bm} 

\usepackage{caption}
\usepackage{subcaption}
\usepackage{float}
\usepackage{array}

\newcommand{\ma}[1]{{\color{black}{#1}}}
\newcommand{\marev}[1]{{\color{black}{#1}}}

\newcommand{\tblue}[1]{{\colorbox{ProcessBlue}{#1}}}
\newcommand{\tyellow}[1]{{\colorbox{yellow}{#1}}}
\newcommand{\torange}[1]{{\colorbox{Apricot}{#1}}}
\newcommand{\tgreen}[1]{{\colorbox{green}{#1}}}

\def\keyFont{\fontsize{8}{11}\helveticabold }
\def\firstAuthorLast{Aktukmak {et~al.}} 
\def\Authors{Mehmet Aktukmak\,$^{1}$, Zeyu Sun\,$^{1}$, Monica Bobra\,$^{2}$, Tamas Gombosi\,$^{3}$, 
Ward B. Manchester\,$^{4}$, Yang Chen\,$^{5}$ and Alfred Hero\,$^{1,5,*}$}
\def\Address{$^{1}$Department of Electrical and Computer Engineering, University of Michigan, Ann Arbor, MI, USA\\
$^{2}$W. W. Hansen Experimental Physics Laboratory, Stanford University, Stanford, CA, USA\\
$^{3}$Department of Aerospace Engineering, University of Michigan, Ann Arbor, MI, USA\\
$^{4}$Department of Climate and Space Sciences and Engineering, University of Michigan, Ann Arbor, MI, USA\\
$^{5}$Department of Statistics, University of Michigan, Ann Arbor, MI, USA}
\def\corrAuthor{Alfred Hero}
\def\corrEmail{hero@umich.edu}

\begin{document}
\onecolumn
\firstpage{1}

\title[Incorporating Polar Field Data for Improved Solar Flare Prediction]{Incorporating Polar Field Data for Improved Solar Flare Prediction}

\author[\firstAuthorLast ]{\Authors} 
\address{} 
\correspondance{} 
\extraAuth{}
\maketitle

\begin{abstract}
In this paper, we consider incorporating data associated with the sun's north and south polar field strengths to improve solar flare prediction performance using machine learning models. \ma{When used to supplement} local data from active regions on the photospheric magnetic field of the sun, the polar field data provides \ma{global information to the predictor}. While such global features have been previously proposed for predicting the next solar cycle's intensity, \ma{in this paper we propose using them to help} classify individual solar flares. \ma{We conduct experiments using HMI data employing four different machine learning algorithms that can exploit polar field information. Additionally, we propose a novel probabilistic mixture of experts model that can simply and effectively incorporate polar field data and provide on-par prediction performance with state-of-the-art solar flare prediction algorithms such as the Recurrent Neural Network (RNN). Our experimental results indicate the usefulness of the polar field data for solar flare prediction, which can improve Heidke Skill Score (HSS2) by as much as 10.1\%} \footnote{{\bf Acknowledgements}: The authors would like to thank Graham Barnes for his valuable comments and suggestions. This work was partially supported by NASA DRIVE Science Center grant 80NSSC20K0600 and W. Manchester was partially supported by NASA grant 80NSSC18K1208.}. 

\tiny \keyFont{ \section{Keywords:} solar flare prediction, polar fields, solar cycle, mixture modeling} 
\end{abstract}

\section{Introduction} \label{sec:intro}
Solar flares rapidly release magnetic energy stored in the Sun's corona, which heats large areas of the surrounding atmosphere. The terrestrial impacts of flares can be serious, especially when they are high-intensity. \ma{ Although weak flares have a limited impact on mankind, strong flares can disturb satellite communication and GPS-navigation systems,  in addition to causing dangerous ionizing radiation exposure to astronauts orbiting the earth.}  
Hence, timely prediction of solar flares, especially the strong ones, is desirable so that precautions can be taken. However, strong flares occur rarely as compared to weak flares, and the underlying mechanisms and the factors that trigger and determine the strength of the flares are not well understood. This makes the prediction of solar flares a complex task.

\ma{As the coronal field is anchored in the photosphere,  the onset of flares has been linked to changes in the photospheric magnetic field \citep{wang2020predicting, SunX:2012}. Consequently, it is reasonable to expect that photospheric magnetic fields in active regions (AR) may be useful indicators of flare-triggering mechanisms. The Helioseismic and Magnetic Imager (HMI) \citep{Scherreretal2012,Hoeksemaetal2014} on Solar Dynamics Observatory (SDO) \citep{PesnellThompsonChamberlin2012} spacecraft has observed the photospheric magnetic field of the sun since 2010, taking high-resolution images with 1 arcsec spatial resolution and 720s cadence. From these magnetic field images,  useful features are computed, known as Space-weather HMI Active Region Patches (SHARP) parameters, which are representative of the magnetic field's scale, energy and complexity \citep{bobra2014helioseismic}. This time-series data has been successfully used for solar flare prediction for Solar Cycle 24  using machine learning algorithms  \citep{chen2019identifying, jonas2018flare,bobra2015solar, nishizuka2017solar, florios2018forecasting, liu2017predicting, camporeale2019challenge, korsos2021testing, campi2019feature}. See also \citep{leka2019comparison} for a comprehensive review.} 

Several supervised machine learning models have been proposed to predict solar flares from HMI observations. NOAA Geo-stationary Operational Environmental Satellites (GOES) flare list \citep{garcia1994temperature} provides peak energy levels for each flare on a logarithmic scale separated into five flare intensity classes, labeled A, B, C, M, X, with A the weakest and X the strongest intensities. These can be used as categorical response variables in machine learning models for flare prediction. Such models use input features, such as HMI SHARP parameters,  to apply off-the-shelf machine learning algorithms to predict the labels of the flares, including discriminant analysis \citep{Lekaetal2018}, linear regression \citep{jonas2018flare}, support vector machine \citep{bobra2015solar, nishizuka2017solar, florios2018forecasting}, k-nearest neighbor \citep{nishizuka2017solar}, random forests \citep{liu2017predicting, florios2018forecasting}, multilayer perceptrons \citep{florios2018forecasting}, and recurrent neural networks \citep{chen2019identifying}. \ma{ Recently, the magnetogram images of the active regions have been used as features in  deep learning models such as convolutional neural networks (CNN) [\citep{huang2018deep}, for prediction of flare intensity class.} Ensemble models have also shown promise as a way of combining predictions of multiple models that use different data sources \citep{Guerraetal2020,sun2022predicting}.

\begin{figure}[h!]
    \centering
    \begin{minipage}[b]{0.8\textwidth}
        \includegraphics[width=\textwidth]{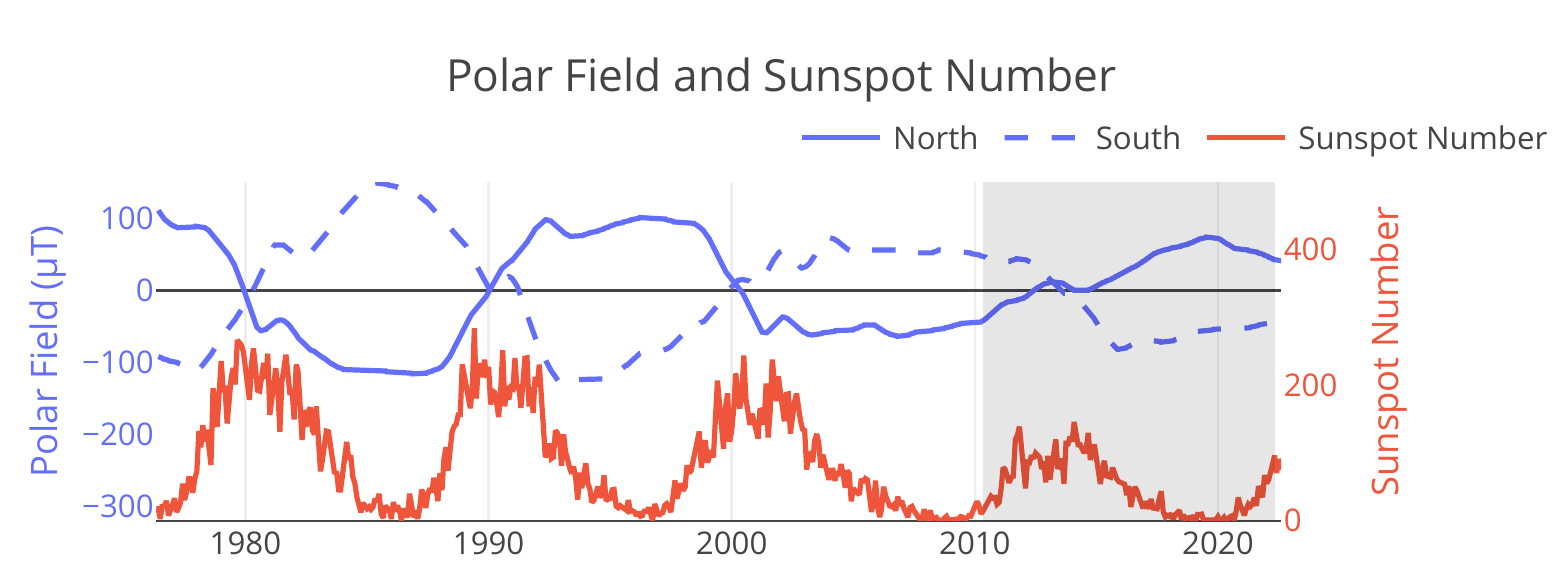}
    \end{minipage}
    
    \begin{minipage}[b]{0.8\textwidth}
        \vspace{1em}
        \includegraphics[width=\textwidth]{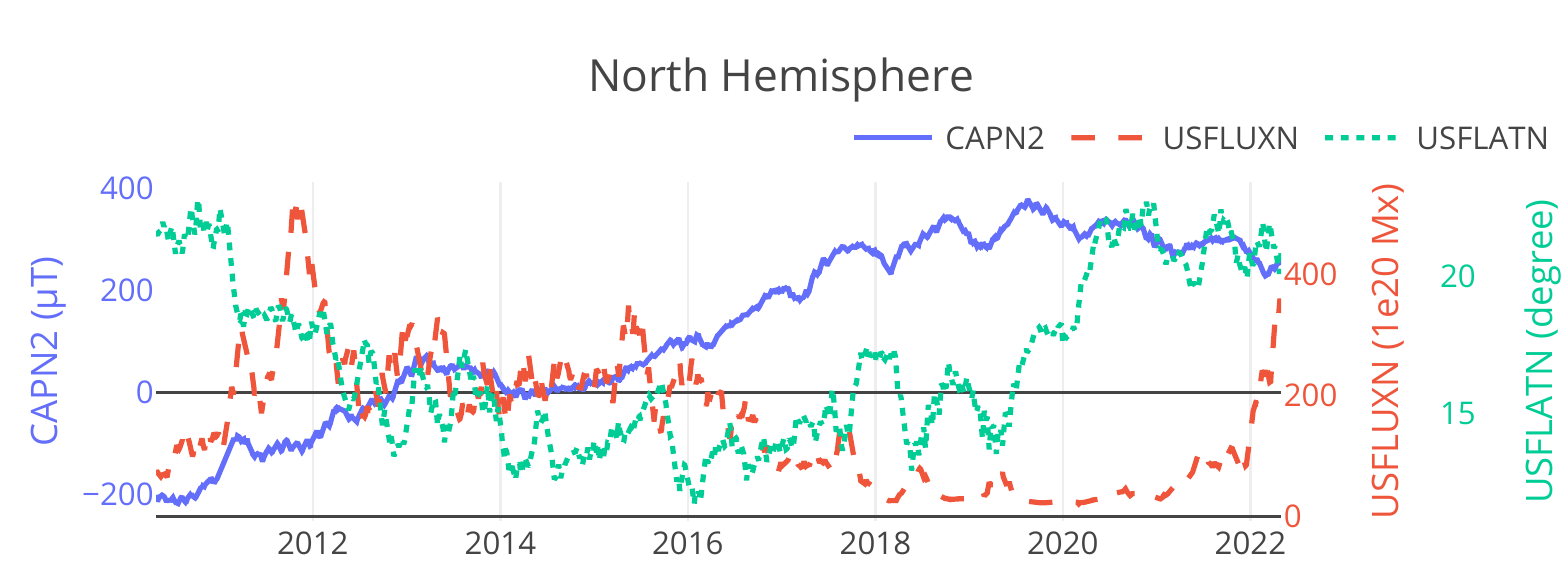}
    \end{minipage}
    
    \begin{minipage}[b]{0.8\textwidth}
        \vspace{1em}
        \includegraphics[width=\textwidth]{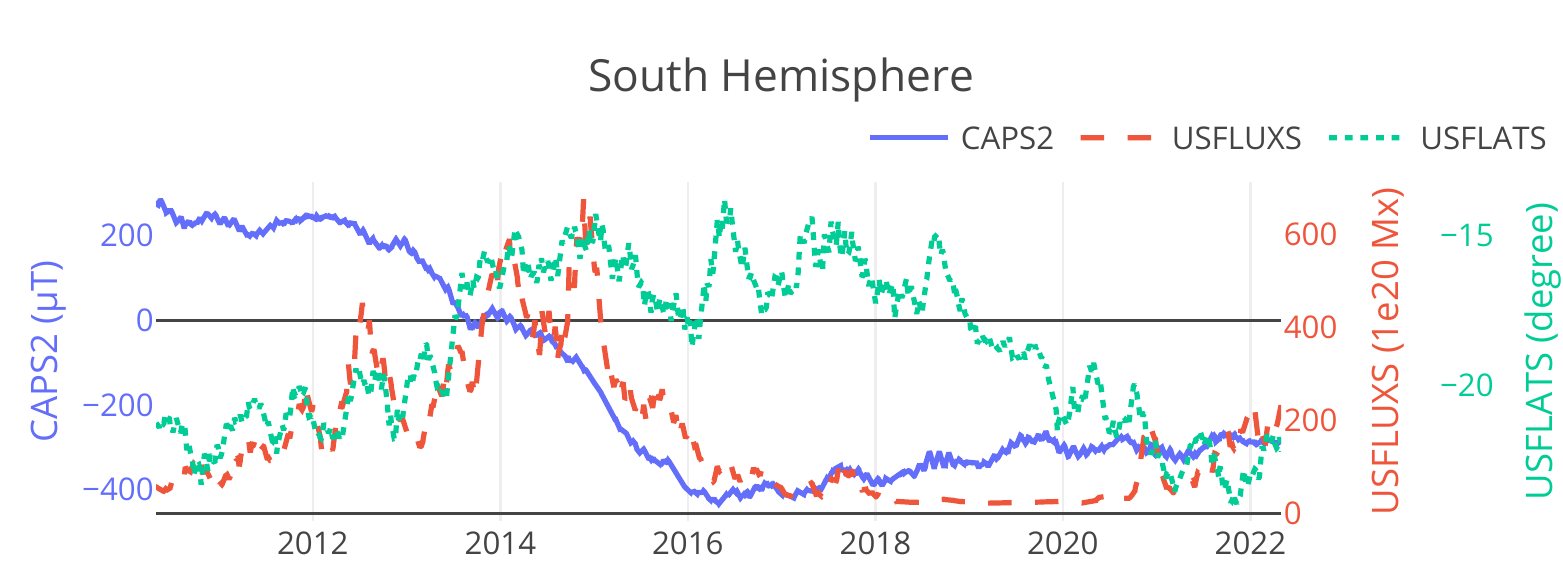}
    \end{minipage}
    \caption{
    \ma{
    \emph{Top}: The north and south polar strength and the monthly mean sunspot numbers plotted over 4 solar cycles. The shaded period is the time during which the HMI instrument has been generating data. Polar fields data are from the Wilcox Solar Observatory \citep{svalgaard1978strength,hoeksema1995large}. Sunspot number data are from WDC-SILSO, Royal Observatory of Belgium, Brussels.
    \emph{Middle}: Solar cycle variation of the north polar field strength (CAPN2), AR unsigned flux (USFLUXN), and flux-weighted AR latitude (USFLATN) obtained from HMI data.
    The unsigned flux in the northern hemisphere (USFLUXN) is positively correlated with the north polar fields (CAPN2), suggesting the inclusion of polar field data as additional features in a machine learning flare prediction model can potentially improve prediction accuracy.
    \emph{Bottom}: Similar to the Middle panel but for the Southern Hemisphere of the sun.
    }}
    \label{fig:polar_fields}
\end{figure}

\ma{Hale's Polarity laws \citep{hale1919magnetic} for sunspots apprise that the polarities of the sunspots (distinguished in pairs with a leading/following counterpart from east to west) have a strong tendency to have the opposite pattern in the northern and southern hemispheres of the sun in a particular cycle. Moreover, the polarities of the spots of the present cycle are the opposite of the polarities in the previous cycle. Additionally, polar field reversals occur systematically at approximately the time of the solar cycle maximum \citep{babcock1959sun}, with peak polar field values occurring at  the time of the sunspot cycle minimum. Thus, the polar fields are 180 degrees out of phase with the sunspot cycle. See Fig.  \ref{fig:polar_fields} for plots over time of the polar fields along with curves of sunspot number, unsigned AR flux, and AR latitude. Numerous works show how the polar field is reformed each cycle  from a small fraction of active region flux, which is transported by meridional flows from mid-latitudes to the poles \citep{hathaway2015solar, wang2005modeling, morgan2017global}.}

It has been demonstrated that the strength of the sun's polar fields can be used to predict the next solar cycle's sunspot intensity based on the understanding of the sun's magnetic dynamo \citep{upton2013predicting, dikpati2006simulating, hiremath2008prediction} \footnote{https://ntrs.nasa.gov/api/citations/20130013068/downloads/20130013068.pdf}.
Weak polar fields at the end of Cycle 23 motivated the authors of the 2005 paper \citep{svalgaard2005sunspot} to forecast a weak Solar Cycle 24. Indeed, Solar Cycle 24 has been the weakest solar cycle in a century. To graphically illustrate the strong association between polar fields and sunspot activity, we the polar fields and monthly mean sunspot number over four solar cycles in the top panel of Fig.  \ref{fig:polar_fields}. It can be observed the magnitude of the difference between the two polar fields at the solar minimum is a good indicator of the intensity of the subsequent cycle.

\ma{
The relation between polar fields and AR magnetic fields is more intricate, involving complex physical mechanisms. This is better observed with higher temporal resolution and in an AR-localized manner. The middle and bottom panels of Fig.  \ref{fig:polar_fields} shows how polar fields interact with total unsigned flux and latitudinal distribution of ARs in the northern and the southern hemisphere, respectively, during the period that HMI has been in operation (the shaded area in the top panel of Fig.  \ref{fig:polar_fields}).
The data are collected from the data product \texttt{hmi.meanpf\_720s}, available at the Joint Science Operations Center (JSOC).
The north and south polar field strengths are represented by CAPN2 and CAPS2, respectively, which correspond to the mean radial field within latitudes N60-N90 and the mean radial field within latitudes S60-S90. The unsigned AR flux can be well approximated by the unsigned flux in between S40 and N40, since flux is concentrated in ARs which are mostly located in those latitudes. Specifically, for the northern hemisphere, we compute the unsigned AR flux (USFLUXN) and unsigned-flux-weighted latitude (USFLATN) via
\begin{align}
    \textrm{USFLUXN} &= \textrm{FLUXN\_P} - \textrm{FLUXN\_N}\,, \\
    \textrm{USFLATN} &= \frac{\textrm{FLATN\_P} \times \textrm{FLUXN\_P} - \textrm{FLATN\_N} \times \textrm{FLUXN\_N}}{\textrm{FLUXN\_P} - \textrm{FLUXN\_N}}\,,
\end{align}
where FLUXN\_P (FLUXN\_N) is positive (negative) flux between N0-N40 above a fixed threshold, and FLATN\_P (FLATN\_N) is positive (negative) flux weighted latitude between N0-N40. Similar quantities are computed for the southern hemisphere as well.
\marev{The active regions' locations follow the well-observed butterfly diagram of sunspots, starting approximately $\pm 40$ degrees latitude at the beginning of a solar cycle, and gradually drifting towards the equator as the cycle progresses. At the cycle minimum, there is an overlap of the old (near the equator) and new (near $\pm 40$ degrees) cycle spots.
These trends can be observed from the lower two panels of Fig.  \ref{fig:polar_fields}.
It is worth noting that the rapid transition in AR latitude at the depth of solar minimum provides a strong signature of the solar cycle when the total flux and polar fields show the lowest variation with time. These strong and complex relations between polar fields and AR fields suggest supplementing SHARP features with HMI polar field strength may improve the prediction of AR-localized flare intensity levels. This is our main motivator for applying machine learning flare prediction approaches to the combined data.
}}


In summary, we validate the usefulness of polar field features in predicting solar flares. The motivation is to assess whether combining SHARP keywords with polar field features can improve solar flare prediction performance or not. To this end, we design an experiment in which there are multiple possible ways of incorporating polar field data, four different machine learning models, and different train-test-validation splitting methods for each algorithm. In addition, we design a novel mixture of experts model tailored for incorporating polar field data for solar flare prediction context. Our experimental results indicate that polar field data can improve solar flare prediction performance. They also show that mixture of experts modeling is a simple yet effective alternative for combining polar field data and SHARP parameters, which can give similar or better performance to much more complicated neural network-based algorithms. 

The rest of this paper is organized as follows. In Section 2, we describe how we incorporate multiple data sources into flare prediction algorithms, and we introduce several data-splitting methods for assessing the performances of the models. In Section 3, we formally define the machine learning models that are compared in this study including the new mixture of experts model. In Section 4, we describe the experimental setup in detail and present our results. 

\section{Data Preparation}


There are three distinct data sources we exploit in this study, namely, SHARP, GOES, and SPF datasets.
The SHARP dataset provides parameters computed from the magnetograms produced by the HMI (See Table     \ref{tab:datasetsapp}) instrument. These parameters are associated with the active regions that occurred from 2010 to 2020. We use the data product \texttt{hmi.sharp\_cea\_720s}, which contains definitive data in Cylindrical Equal-Area (CEA) coordinates and is available at the Joint Science Operations Center (JSOC).  
The GOES dataset includes solar flare event records measured by Geostationary Operational Environmental Satellites (GOES). This dataset can be downloaded using the SunPy package \footnote{\url{https://sunpy.org/}} and it covers the HMI time period from 2010 to 2020. The flare events are labeled by A, B, C, M, and X, based on the peak intensity of X-ray flux measures. 
The Solar Polar Fields (SPF) dataset consists of two solar polar field features, namely, CAPS2 and CAPN2. These features correspond to the mean radial field in latitudes N60-N90, and the mean radial field in latitudes S60-S90, respectively. Both HMI and MDI instruments record these features, which can also be downloaded from JSOC. \ma{The MDI instrument was operational from 1996 to 2010 and the HMI has been operational since 2010.}

\begin{table}[]
    \centering
    \small
    \begin{tabular}{m{1.3cm} |m{1.3cm} | m{1.3cm} | m{1.3cm} | m{5cm}}
        \hline
        Dataset & \# of instances & B-class ratio & MX-class ratio & Parameters \\
        \hline
        SHARP & 2663 & 0.49 & 0.51 & USFLUXL, MEANGBL, RVALUE, AREA, TOTUSJH, TOTUSJZ, SAVNCPP, ABSNJZH, TOTPOT, SIZEACR, NACR, MEANPOT, SIZE, MEANJZH, SHRGT45, MEANSHR, MEANJZD, MEANALP, MEANGBT, MEANGAM, MEANGBH, NPIX\\
        \hline
    \end{tabular}
    \caption{SHARP dataset statistics after preprocessing step. See Table 3 in the appendix for the definitions of the keywords.}
        \label{tab:datasets}
\end{table}

\subsection{SPF Preprocessing}

The noise levels on CAPS2 and CAPN2 features are significant, and the noise level is different in the HMI and MDI instruments. Hence, we consider smoothing by using a Kalman filter to filter out noise over these features. \ma{Note that the variance over the time series data is not solely due to noise. There is strong yearly modulation due to the Earth-solar B0 angle, where the solar rotation axis varies by $\pm$7 degrees. Furthermore, there is strong 27-day modulation due to solar rotation.}

\begin{figure} 
    \centering
    \includegraphics[width=1\textwidth]{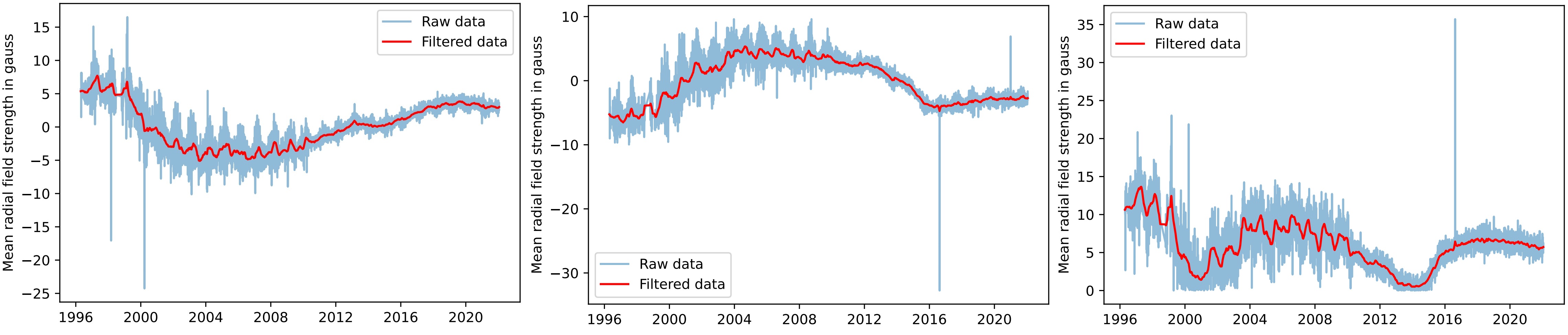}
    \caption{Raw (blue curve) and Kalman filtered (red curve) polar field features.  Polar field strengths at the north pole CAPN2, south pole CAPS2, and the solar cycle approximation, defined as the unsigned polar field difference $|CAPS2(t) - CAPN2(t)|$, are shown in the left, middle, and right-hand panels respectively. The data source switched from MDI to HMI in 2010, which is accompanied by a visually obvious reduction in noise. 
    The HMI data source has lower noise due to its high resolution. 
    }
    \label{fig:polar_field_features}
\end{figure}

\ma{We implement the Kalman filter using the local level model \citep{murphy2012machine} where the state transition and observation models are defined as follows:
\begin{equation}
    a_t = a_{t-1} + \epsilon_t^a,
\end{equation}
\begin{equation}
    y_t = a_t + \epsilon_t^y,
\end{equation}
where $y_t$ is the raw noisy observation at time $t$, $a_t$ is the filtered data, and $\epsilon_t^y \sim \mathcal{N}(0,\lambda_y)$ and $\epsilon_t^a \sim N(0, \lambda_a)$ are the noise processes over the observed and hidden spaces with variances $\lambda_y$ and $\lambda_a$, respectively. By adjusting these levels instrument-wise, we perform smoothing for the data of both instruments. See Fig. \ref{fig:polar_field_features} for the raw and filtered features.} 


\subsection{Observation Preprocessing}

In this section, we describe how we use the HMI SHARP dataset  and the GOES flare intensity labels to construct the training and test data for our discriminative machine learning algorithms, developed in the next section. Our Preprocessing steps for these data follow the approach in \citep{jiao2020solar, chen2019identifying, sun2022predicting}. 

We only include ARs having unique NOAA AR numbers \citep{jiao2020solar} into our training set.  We eliminate low-quality observations by excluding ARs that occur outside of range $\pm70$ relative to the central meridian \citep{chen2019identifying}.
\ma{Similarly to \citep{bobra2014helioseismic,chen2019identifying} we focus on predicting whether a solar flare from a given active region will be strong (M/X class) or weak (B class). The A-class flares are excluded as they are almost indistinguishable from the background radiation. The C-class flares are frequently missing their NOAA AR numbers in the GOES database, making their locations difficult to track [3].  Exclusion of A and C flares and combining M and X class flares together creates a binary classification problem  that enables the comprehensive comparative computational analysis we report in Sec. 4. This is a step along the way of establishing the value of polar field data for the  more complex classification task of classifying between the five labels A, B, C, M, X, which is worthwhile of future study. }



\ma{As in \citep{jiao2020solar}, the class label of an AR is designated as the strongest flare occurring  in a 24-hour prediction period, according to the GOES database. Thus an AR is assigned a B label if there are no M/X flares within 24 hours, while it is assigned label M/X otherwise.  This is accomplished as follows. For each of the flares occurring in an AR,  variables are extracted from the time segment spanning a 24-hour time period from 48 hours to 24 hours before the occurrence of the flare. Since the cadence of the HMI observations is one sample every 12 minutes this yields a matrix of 22 SHARP predictor variables defined over 120-time points. If, for a given flare, the corresponding predictor variables have NaN values no class prediction is generated and we omit the flare from our analysis. Furthermore, in our analysis, we do not use any flare labels in the first 24 hours of the beginning of an ARs since such flares would have no associated observed predictor variables.
Once these omissions are taken into account, the total number of observations (predictor variables-label pairs) is 2663. See Table \ref{tab:datasets} for a summary of our set of SHARP features. }

\subsection{Train-Validation-Test Splitting}

\ma{We consider three different methods for splitting the set of AR samples into train-test sets. In all three methods, we split the AR's into 80\% train and 20\% test samples.
i) Instance-based splitting. The indexes of the splits are generated randomly without replacement.  This split can measure the goodness-of-fit but is prone to information leakage when the observations of a particular AR are scattered among the splits \citep{sun2022predicting}. 
ii) AR-based splitting. AR numbers are used instead of the observation indexes for random index generation.  This split prevents the information leakage that happens in Random splitting but may introduce another form of information leakage due to sympathetic flaring between neighboring ARs \citep{sun2022predicting}.
iii) Month-based splitting. The timestamps are used to split the sets in terms of months with the same ratio. This split can also measure the goodness of fit and is closer to the operational forecasting setting. 
Each feature in the dataset is normalized over the time windows of the training data. Specifically, for each AR the mean and standard deviation of each feature is computed over the time points of the training sample.   Then for each AR and each variable, the mean is subtracted and the result is divided by the standard deviation for both training and test samples.} 


\section{Prediction algorithms}
\label{sec:algorithms}

We consider four machine learning models that can exploit polar field features for the flare prediction task: i) Logistic Regression (LR); 
ii) Mixture of Experts (MOE); 
iii) Multi-Layer Perceptron; 
iv) Recurrent Neural Network. 

\subsection{Logistic Regression}

Logistic Regression (LR) is a model for which the conditional distribution of the response variable as a Bernoulli distribution conditioned on the features of an AR sample in the case of binary classes, and categorical (multinomial) distribution for multiple classes.   Specifically, for the binary case, LR models the conditional distribution of the response variable  as
\begin{equation}
	p(y_{i} | \bm{x}_{i}, \Theta) = Ber(y_{i} | \sigma (\Theta^T \bm{x}_i) ),
\end{equation}
where $i$ indexes the data points, $y_{i} = \{0, 1\}$ denotes the response variable (i.e., flare class), and $\bm{x}_i$ corresponds to the SHARP parameters. The model parameters $\Theta$ are estimated from the training set by using Maximum Likelihood Estimation (MLE). In particular, the MLE objective over the $N$ AR samples $\{(\bm{x}_i,y_i)\}_{i=1}^n$ in the training set: 
\begin{equation} 
	\ell(\Theta) = \sum_{i=1}^N  \log p( y_i | \bm{x}_i, \Theta),
	\label{eq:obj_lr}
\end{equation}
which is convex in $\Theta$, and has a unique solution. We use LBFG-S \citep{liu1989limited} to optimize the model parameters. In order to introduce polar field features to LR, a simple approach is to concatenate them with the SHARP parameters and apply the algorithm with these augmented features. 

\subsection{Mixture of Logistic Regressions}
A richer family of conditional distributions can be obtained by using a mixture of experts (MOE) framework \citep{yuksel2012twenty}. For classification tasks, each expert is an LR model. The predictions of the experts are mixed probabilistically according to mixture coefficients, which are obtained by using another linear model called the gating function. The output of the gating function is constrained to be a simplex to support proper mixing coefficients. This is achieved by using a Softmax function $\mathcal{S}$, which is applied after the linear transformation of the input features of the gating function. We propose to use polar field features as the input to the gating function. The main intuition is that the experts will be forced to focus on specific portions of the solar cycle. 

Let $\bm{z}_{i}$ denote a $K$ dimensional categorical valued latent variable that indicates which expert is active for data point (AR) $i$, where $K$ is the total number of experts initialized. The categorical conditional distribution of the latent variable $\bm{z}_{i}$ induced by the gating function is given as
\begin{equation}
	p(z_{i} | \bm{t}_{i}, \bm{V}) = Cat(z_{i} | \mathcal{S} (\bm{V}^T \bm{t}_i) ),
\end{equation}
where $\bm{t}_i$ denotes the polar field features associated with data point $i$, and $\bm{V}$ represents the gating function parameters that map features to the logits of the categorical distribution. After the softmax function, we obtain proper mixture coefficients for data point $i$. Since the experts are logistic regression models, their conditional distributions are given as 
\begin{equation}
	p(\bm{y}_{i} | \bm{x}_{i}, z_{i} = k , \Theta_k) = Ber(y_{i} | \sigma (\Theta_k^T \bm{x}_i) ),
\end{equation}
where $\Theta_k$ denotes the parameters of the expert $k$. The marginal likelihood of this model can be computed by integrating out the latent variables as follows
\begin{equation} \label{eq:mix_obj}
	p(\bm{y}_{i} | \bm{x}_{i}, t_i) = \sum_{k=1}^K p(z_i = k | t_i, \bm{V} ) p( y_i | \bm{x}_i, z_i = k, \Theta_k).
\end{equation} 

The objective is to maximize marginal log-likelihood with respect to the logistic regression and gating function parameters. The Expectation Maximization algorithm is a suitable optimization method for latent variable models. The algorithm optimizes the expected complete data log-likelihood (ECLL) instead of marginal, where the former forms a lower bound for the latter. There are two alternating steps, E and M steps. In the E-step of the algorithm, the posterior distributions of the latent variables are computed and their expectations are plugged into the complete data log-likelihood to obtain ECLL. Subsequently, the ECLL is maximized with respect to the model parameters in the M-step. These steps are iterated until ECLL converges. For the mixture of experts model, the ECLL is given as 
\begin{equation} \label{eq:obj_em}
	\sum_{i=1}^N \sum_{k=K}^K r_{ik} \log [ \mathcal{S}_k ( \bm{V}^T t_i ) p ( y_i | \bm{x}_i, \Theta_k)],
\end{equation}
where we replace the explicit expression of the prior distribution of mixture coefficients. $r_{ik}$ denotes the posterior distribution of $z_i$, which can be computed as 
\begin{equation} \label{eq:e_step}
	r_{ik} = p (z_i = k | \bm{x}_{i}, y_{i}, t_{i}, \Theta_k) \propto \mathcal{S} ( \bm{V}^T t_i ) p( y_i | \bm{x}_i, z_i = k, \Theta_k),
\end{equation}
in the E-step of the EM algorithm. Replacing the posterior in the complete data log-likelihood, we get parameter-specific objectives to optimize in the M-step. Specifically, for the model parameters of the experts, the objective becomes
\begin{equation} \label{eq:m_step_Th}
	\ell(\Theta_k) = \sum_{i=1}^N  r_{ik} \log p( y_i | \bm{x}_i, z_i = k, \Theta_k),
\end{equation}
which can be computed by using any gradient-based optimizer. Note the similarity of this objective with Eq. \ref{eq:obj_lr}. Hence, the former corresponds to weighted Logistic Regression. For the parameters of the gating function, the objective takes the following form
\begin{equation}\label{eq:m_step_V}
	\ell(\bm{V}) = \sum_{i=1}^N \sum_{k=1}^K  r_{ik} \log \mathcal{S}_k ( \bm{V}^T t_i ),
\end{equation}
which is also similar to the Logistic Regression objective in Eq. \ref{eq:obj_lr} except that the response variable is simplex instead of discrete. This objective can also be maximized with any gradient-based optimizer. We use LBFG-S for all gradient-based optimization tasks. In summary, the E-step constitutes computing posterior distributions of the latent variables given the current values of the parameters by using Eq. \ref{eq:e_step}, the M-step constitutes maximizing Eq. \ref{eq:m_step_Th} and \ref{eq:m_step_V} with respect to $\Theta_k$ and $\bm{V}$. When Eq. \ref{eq:obj_em} converges, the iterations are halted. If we don't use polar field features, mixing coefficients are unconditional as analogous to Gaussian Mixture Models.

\subsection{Multi-layer Perceptron}

The Multi-layer Perceptron (MLP) is a non-linear discriminative classification model. MLPs can be more powerful than their linear counterparts when the interactions between the features and the response variables can not be explained by linear decision boundaries. An MLP can be obtained by stacking multiple fully connected layers. Each layer performs an affine transformation, and subsequently applies a deterministic non-linear function, e.g., softmax, relu, tanh. Stacking multiple layers result in a complex deterministic function of the input features. Let $f_{\Theta}$ denote this function with the exception that the last layer does not apply any activation functions. Here, $\Theta$ represents the set of the affine transformation parameters of all the layers. Then, the conditional distribution for a binary classification task is given as
\begin{equation}
	p(y_{i} | \bm{x}_{i}, \Theta) = Ber(y_{i} | \sigma (f_{\Theta}(\bm{x}_i)),
\end{equation}
where the loss function is the same with LR as in Eq. \ref{eq:obj_lr}.

The maximization of this distribution can be achieved by using a back-propagation algorithm \citep{hecht1992theory}. We consider using polar field features as the additional features by concatenating them with the SHARP parameters. 

Analogously to LR, a mixture of MLPs can be constructed by initializing multiple MLPs and mixing them using the probabilistic objective given in Eq. \ref{eq:mix_obj} \citep{bishop1994mixture}. However, in the MLP case, the EM algorithm is no longer applicable. Fortunately, back-propagation can still be used to maximize the objective. Similar to the Mixture of Logistic regression, a separate gating function can be initialized that maps polar field features to the mixture coefficients. This function can be linear or non-linear (MLP). In our experiments, we did not observe a significant performance change for Mixture of MLPs over NN-Augmented. Hence, we omit the Mixture of MLPs in this study. 

\subsection{Recurrent Neural Network}
The aforementioned algorithms do not take the sequential nature of the observations into account. A more principled approach would model temporal data by assuming Markov chains between the consecutive inputs or outputs. Recurrent Neural Networks (RNN) do that by forming a first-order Markov chain between the hidden units. RNNs have shown significant success for sequential data modeling, and LSTM, a variant of RNNs, achieves state-of-the-art performance in flare forecasting from time series SHARP data \citep{chen2019identifying, liu2019predicting,sun2022predicting}. For this study, we use a particular version to classify the sequences of the input observations, also known as time series classification. The conditional distribution of the model is given as:
\begin{equation}
	p(y_{i} | \bm{x}_{i,t=1:T}, \Theta) = Ber(y_{i} | \sigma (f_{\Theta}(\bm{x}_{i,t=1:T})),
\end{equation}
which implies that the observed label is dependent on a sequence of input data points. The length of the sequence is denoted as $T$ and is a hyper-parameter of the model. 
The log-likelihood function associated with the RNN model is 
\begin{equation} 
	\ell(\Theta) = \sum_{i=1}^N  \log p( y_i | \bm{x}_{i,t=1:T}, \Theta),
\end{equation}
where $\Theta$ represents the parameters of the model and can be optimized by using the backpropagation algorithm and $T=120$ is the number of temporal samples in each AR training sample.  The polar field features are incorporated as additional input features by concatenating them with the SHARP parameters. 

\section{Experiments}

\subsection{Configurations}
The 4 algorithms described in Sec. \ref{sec:algorithms} were compared in terms of relative performance improvements when adding the polar field to the SHARP features.  Our SHARP dataset consists of 2663 samples with 22 parameters over 120 time points. As described above we consider 3 different data splits: instance-based, AR-based, and month-based. For polar field features, we create two modes: raw CAPS2 and CAPN2 used as two separate input features; and cycle, where the absolute difference of CAPS2 and CAPN2 is used as a lower dimensional solar cycle approximation (see Fig. \ref{fig:acc_solar_cycle}). We also consider smoothed versions of CAPS2 and CAPN2 by using the Kalman filter, and unfiltered versions of these features. Hence, the total number of experimental configurations is 12 if we don't use the polar field feature, and 48 otherwise. We perform cross-validation by using training splits for each configuration to select the regularization strengths of the algorithms. Also, we perform 20 runs to assess the significance of the accuracies. Hence, the total number of runs is around 7k. 
 
\subsection{Tuning}
The hyper-parameters of the machine learning algorithms must be tuned for fair performance comparison. Cross-validation on the training set is used to select the best hyper-parameters for each experimental configuration. First, the training dataset is divided into 5 folds. The first fold is held out for measuring the performance of a hyper-parameter setting while the rest are used for training the algorithm. Then the hold-out set is alternated one by one so the total number of empirical realizations of performance for each hyper-parameter setting is 5. The average of these 5 values is then used as the score of that particular setting. We use the Heidke Skill Score HSS2 \citep{cohen1960coefficient, bobra2015solar} for quantifying performance, which is defined as follows:
\begin{equation}
    HSS2 = \frac{2 \times [ (TP \times TN) - (FN \times FP)]}{(TP + FN)(FN + TN) + (TP + FP)(FP+TN)},
\end{equation}
\ma{where TP, TN, FP, and FN correspond to true-positive, true-negative, false-positive, and false-negative values computed from binary classification predictions and ground truth, respectively.} 

All the algorithms studied were implemented with the standard L2-norm regularization approach to minimize overfitting. Regularization is accomplished by a penalty method, where the sum of the L2-norm squared of each parameter vector is weighted by a regularization parameter and added to the log-likelihood functions of each of the predictor models discussed in Sec. \ref{sec:algorithms}. The regularization parameters are tuned to optimize the performance of each algorithm. For LR, MLP, and RNN, the regularization strength is a single scalar that is common for all the elements of the parameters. The grid search range for this scalar is chosen as $[0.1, 1, 10]$. For MOE, we have separate parameters for the gating function and the experts. We use one scalar regularization parameter for the parameters of the gating function, and another for the parameters of the experts. The grid range for these two hyper-parameters is similarly determined as $[0.1, 1, 10]$. 

In addition to regularization tuning, the mixture model has a hyper-parameter associated with the total number of experts being initialized. The grid search range for this hyper-parameter is set as $[2, 3, 4, 5, 6, 7, 8]$. We fix the network structure of RNN and MLP. In particular, 32 nodes with one hidden layer are initialized for both models.

\subsection{Significance Assessment}
To assess the significance of the results, we repeat each experimental configuration $n=20$ times with and without using polar field data. We then compute p-values from one-sided paired t-tests. The t-statistic is computed as $t = \Bar{d}\sqrt{n}/s$, where $\Bar{d} = \frac{1}{n}\sum_{i=1}^n d_i$ is the mean of the difference vector $\bm{d}$ whose elements are given by $d_i = y_i - x_i$ where $y_i$ and $x_i$ are HSS2 scores with and without polar field data for trial $i$, respectively. Then, under the Student-t distribution with degrees of freedom $n - 1$, we compute the right-tail p-value. 

\subsection{Results}

\begin{table}[]
    \centering
    \small
        \begin{tabular}{c  c c  c  c  c c c c c}
        \hline
        Algorithm &Filter & Mod & Split & p-value & t-score & hss2(s+p) & hss2(s) & hss2-diff & hss2(p) \\
        \hline
        LR &    \tblue{Yes}     &  \tblue{Raw}  & \tblue{Month} & \tblue{0.0274} & \tblue{2.628} & \tblue{0.7745} & \tblue{0.7155} & \tblue{\%8.2} & \tblue{0.3010}\\
        &\tyellow{No}         &  \tyellow{Cyc}  & \tyellow{Inst} & \tyellow{0.0439} & \tyellow{2.1271} & \tyellow{0.8143} & \tyellow{0.802} & \tyellow{\%1.5} & \tyellow{0.3628}\\
        \hline
        &\tyellow{No}         &  \tyellow{Cyc}  & \tyellow{Inst} & \tyellow{0.0002} & \tyellow{4.3055} & \tyellow{0.8223} & \tyellow{0.7992} & \tyellow{\%2.9} & \tyellow{0.3484}\\
        MOE&\torange{Yes}     &  \torange{Raw}  & \torange{Inst} & \torange{0.0005} & \torange{4.0369} & \torange{0.8254} & \torange{0.792} & \torange{\%4.2} & \torange{0.3867}\\
        &\tgreen{Yes}        &  \tgreen{Cyc}  & \tgreen{Inst} & \tgreen{0.0155} & \tgreen{2.6055} & \tgreen{0.8205} & \tgreen{0.8001} & \tgreen{\%2.5} & \tgreen{0.3689}\\
        &\tblue{Yes}        &  \tblue{Raw}  & \tblue{Month} & \tblue{0.0386} & \tblue{2.4202} & \tblue{0.7502} & \tblue{0.6827}  & \tblue{\%9.9} & \tblue{0.2485}\\
        \hline
        MLP&\torange{Yes}     &  \torange{Raw}  & \torange{Inst} & \torange{0.0000} & \torange{6.8479} & \torange{0.7992} & \torange{0.7896} & \torange{\%4.9} & \torange{0.4384}\\
        &No         &  Cyc  & Month & 0.0481 & 2.2856 & 0.7653 & 0.7149 & \%7.0 & 0.3658\\
        \hline
        &\torange{Yes}        &  \torange{Raw}  & \torange{Inst}     & \torange{0.0002} & \torange{4.4235} & \torange{0.826} & \torange{0.7947} & \torange{\%3.9} & \torange{0.4143}\\
        RNN&Yes     &  Cyc  & Month    & 0.0463 & -2.3092 & 0.6117 & 0.6756 & -\%9.4 & 0.3627\\
        &\tblue{Yes}        &  \tblue{Raw}  & \tblue{Month}    & \tblue{0.0478} & \tblue{2.2888} & \tblue{0.7194} & \tblue{0.6544} & \tblue{\%9.9} & \tblue{0.2791}\\
        &No         &  Raw  & AR       & 0.0221 & 2.4469 & 0.6977 & 0.6337 & \%10.1 & 0.3952\\
        \hline
    \end{tabular}
    \caption{\ma{Table of statistically significant (p-value$\leq$5\%)  performance improvements due to adding polar field features to SHARP features, The hss2 skill score argument (p) means only polar fields data, (s) means only SHARP parameters, and (s+p) means both data sources are used as input to the algorithms. (diff) is the mean hss2 score percentage difference when the polar field is incorporated in addition to SHARP parameters.  The highlight colors in the table specify the common configurations where MOE achieves statistically significant performance improvements due to adding the polar field. The full tables including the statistically non-significant configurations are provided in the supplementary.}}
    \label{tab:res_hss2}
\end{table}

The results are compiled in Table \ref{tab:res_hss2}. We report HSS2 computed on the test set as the performance metric. For each algorithm, we only keep the configurations with statistically significant performance improvement due to the incorporation of polar field data, where significance is defined as a p-value less than 0.05 (5\% level) of a one-sided paired t-test of equal mean performance, computed from 20 runs.
For the LR model, there are two configurations that exhibit significant performance differences when polar field features are incorporated. Both of them imply that performance improves with the incorporation polar field. \ma{Particularly when we use unfiltered cycle approximation for the polar field features and filtered raw features, we obtain performance boost for the month (8.2\%) and instance-based splits (1.5\%), respectively.
For the MOE model, there are four significant cases with all of them yielding performance improvements. The performance boost is up to 9.9\%. The model can exploit the polar field augmented data regardless of filtering and preprocessing options when the splits are instance and month based to improve flare classification performance.
For filtered raw features, the MLP improves
performance on instance-based splitting by 4.9\% and non-filtered cycle approximation on month-based splitting improves by 7.0\%. The RNN performance is boosted regardless of filtering, preprocessing, and splitting options for four significant cases, achieving improvements as high as 10.1\%. In summary, out of 12 significant cases for four algorithms, 11 of them provide performance increase, which suggests that polar field features to have high potential to improve solar flare prediction performance.}

\ma{The column hss2(p) shows the prediction performance of the algorithms when only polar field data is used. The results indicate that sole use of polar field data gives poor skill scores, which is less than half of those of the SHARP data alone, for solar flare prediction. However, the hss2(p)  scores are far from zero since solar cycle intensity and polar fields are correlated, as discussed in Section \ref{sec:intro}. Solar cycle intensity affects the expected intensities of the flares in particular time frames. Due to this connection, one can expect that coarse predictions about the classes of solar flares can be made with polar field data since it captures information about the solar cycle. 
}

\begin{figure}
    \centering
    \scalebox{.8}{\includegraphics{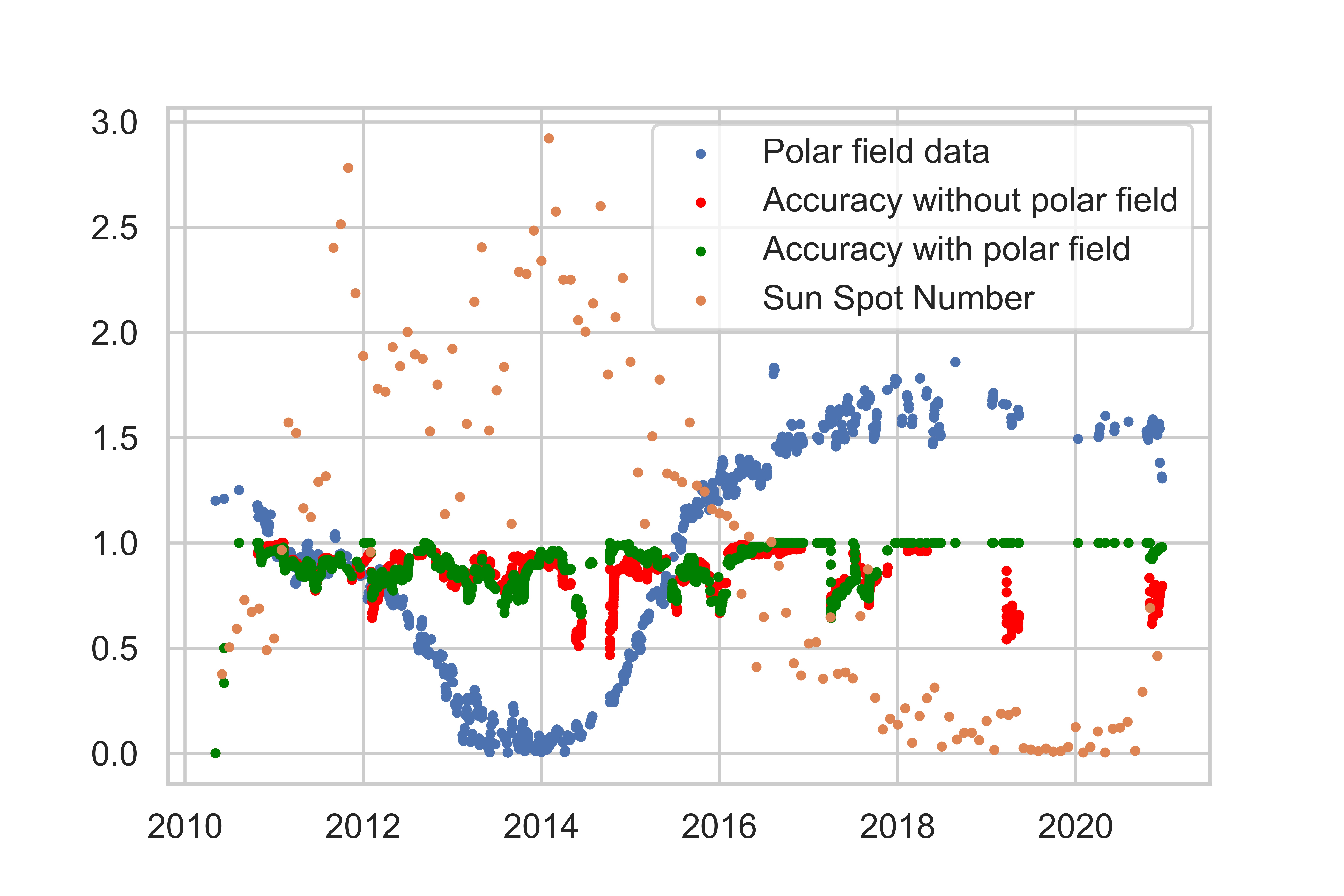}}
    \caption{The classification accuracy curves of MOE with and without using polar fields data w.r.t. to the sunspot number when the splitting method is instance-based. Cycle approximation with filtered data is used as input. The curves are moving averaged over 90 days. The incorporation of the data substantially improves classification accuracy in some portions of the solar cycle. }
    \label{fig:acc_solar_cycle}
\end{figure}

For the MOE model, the solar cycle dependency of the classification accuracy is shown in Fig. \ref{fig:acc_solar_cycle} by repeating five experiments using instance-based splitting. We plot sunspot numbers, polar fields data (cycle approximation), accuracy without polar fields data, and accuracy with polar fields data. The classification accuracy curves are smoothed out with a 90-day moving average. We observe that MOE can exploit polar field data to improve classification accuracy in some parts of the solar cycle. The average classification accuracies over the whole HMI period are 0.88 and 0.91 for without and with polar field data, respectively.
One can also observe the high negative correlation between the sunspot number and the curve of the polar field, which is computed as -0.86.

Finally, we compare the performance improvements of the models in  Table \ref{tab:res_hss2}. Only those configurations for which the p-value was significant at a level of 5\% are shown.
Among all algorithms MOE stands out as the only one to attain highly significant performance improvement (p-value=0.0002) when no Kalman filtering is used, the magnitude difference north and south polar field data are used (the solar cycle approximation), and the splitting is instance based.  MOE, MLP, and RNN, all have highly significant  performance improvement when the polar field data is smoothed using the Kalman filter, the northern and southern polar fields are not combined into a cycle approximation (raw) and the split is instance based. 
Overall, among the five common configurations for which MOE achieves statistically significant improvement at the 5\% level,  MOE exhibits better performance in three of them, comparable performance in one of them, and slightly worse performance in the other one. We conclude that MOE is a promising machine learning model that can incorporate polar field data to improve flare classification accuracy. 

We conclude this section by discussing some of the limitations of this study, which create opportunities for future research. Firstly, while we have established the value of polar field data when the train vs test data split is randomized over time, our conclusions may not translate to operational flare forecasting where train and test data are drawn from past and future observations, respectively. We could not implement such a temporally defined data splitting to train our algorithms due to the fact that HMI data is currently available only over a single solar cycle. Secondly, it is possible that we could strengthen our conclusions by accounting for the temporal and longitudinal correlations of the polar field and active region magnetograms.   In particular, instead of summarizing the polar field as the difference between both north and south polar fields (CAPN2 and CAPS2) we might be able to improve our model by incorporating the spatial coordinates  of the active regions and independently applying CAPN2 and CAPS2 to the northern and southern hemisphere active regions, respectively (see Fig. \ref{fig:polar_fields}).   Alternatively, it might be beneficial to incorporate a dynamic physics-based prediction model to account for dependencies between past active region activity and present polar field intensity. Furthermore, the added value of using polar field data may increase if we expanded the feature dimension to use the higher resolution HMI image data instead of just SHARP summarized data of active regions.                       

\section{Conclusions}
In this paper, we investigated the utility of the sun's north and south polar field strength for improving solar flare prediction from HMI data. We presented an extensive experimental study, that analyzes multiple ways of preprocessing schemes for polar field data, several machine learning models that can incorporate this additional data source, diverse experimental conditions, and multiple splitting methods. The mixture of experts model was introduced for this particular global data incorporation task Our experimental results showed that polar field data is valuable for improving classification performance in multiple configurations. Additionally, we demonstrated that the proposed mixture of experts model can effectively supplement   SHARP parameters with polar field data with for improved flare prediction. 

\ma{The results suggest a causal connection between isolated ARs captured in the SHARP data and the global magnetic field captured in the polar field.  The connection could come through the polar field itself due to the causal relationship between the event of decaying ARs and the subsequent drift of enhanced field towards poles. Hence, AR activity is anti-correlated to the polar field with a certain phase shift. In this study, we find that the probability of individual flare events is also correlated to the total sunspot number through the solar cycle, which closely follows the polar field strength. To put this work in context we point out that the polar field data has been used to inform solar cycle predictions with reasonable accuracy \citep{Pesnell:2012, Pesnell:2020}. We believe this work represents the first time the global field has been used to augment the prediction of solar flares. 

In future work, there are several promising directions to enhance current results. The phase shift between the solar cycle/AR activity and polar field strengths can be estimated jointly by developing a more sophisticated model that can estimate and take into account phase shifts to improve prediction performance. The ARs in the northern and southern hemispheres can be modeled separately for more focused predictions. The correlation between the polar field strengths and other global indicators such as total active region area and flux-weighted latitude at a certain time can be investigated and possibly utilized for improved prediction performance. Finally, the data and the model may be further refined 
 by using HMI high-dimensional image data instead of SHARP features.}
 

\bibliographystyle{plainnat}
\bibliography{frontiers.bib}

\begin{thebibliography}{42}
\providecommand{\natexlab}[1]{#1}
\providecommand{\url}[1]{\texttt{#1}}
\expandafter\ifx\csname urlstyle\endcsname\relax
  \providecommand{\doi}[1]{doi: #1}\else
  \providecommand{\doi}{doi: \begingroup \urlstyle{rm}\Url}\fi

\bibitem[Babcock(1959)]{babcock1959sun}
Harold~D Babcock.
\newblock The sun's polar magnetic field.
\newblock \emph{The Astrophysical Journal}, 130:\penalty0 364, 1959.

\bibitem[Bishop(1994)]{bishop1994mixture}
Christopher~M Bishop.
\newblock Mixture density networks.
\newblock 1994.

\bibitem[Bobra and Couvidat(2015)]{bobra2015solar}
Monica~G Bobra and Sebastien Couvidat.
\newblock Solar flare prediction using sdo/hmi vector magnetic field data with
  a machine-learning algorithm.
\newblock \emph{The Astrophysical Journal}, 798\penalty0 (2):\penalty0 135,
  2015.

\bibitem[Bobra et~al.(2014)Bobra, Sun, Hoeksema, Turmon, Liu, Hayashi, Barnes,
  and Leka]{bobra2014helioseismic}
Monica~G Bobra, Xudong Sun, J~Todd Hoeksema, M~Turmon, Yang Liu, Keiji Hayashi,
  Graham Barnes, and KD~Leka.
\newblock The helioseismic and magnetic imager (hmi) vector magnetic field
  pipeline: Sharps--space-weather hmi active region patches.
\newblock \emph{Solar Physics}, 289\penalty0 (9):\penalty0 3549--3578, 2014.

\bibitem[Campi et~al.(2019)Campi, Benvenuto, Massone, Bloomfield, Georgoulis,
  and Piana]{campi2019feature}
Cristina Campi, Federico Benvenuto, Anna~Maria Massone, D~Shaun Bloomfield,
  Manolis~K Georgoulis, and Michele Piana.
\newblock Feature ranking of active region source properties in solar flare
  forecasting and the uncompromised stochasticity of flare occurrence.
\newblock \emph{The Astrophysical Journal}, 883\penalty0 (2):\penalty0 150,
  2019.

\bibitem[Camporeale(2019)]{camporeale2019challenge}
Enrico Camporeale.
\newblock The challenge of machine learning in space weather: Nowcasting and
  forecasting.
\newblock \emph{Space Weather}, 17\penalty0 (8):\penalty0 1166--1207, 2019.

\bibitem[Chen et~al.(2019)Chen, Manchester, Hero, Toth, DuFumier, Zhou, Wang,
  Zhu, Sun, and Gombosi]{chen2019identifying}
Yang Chen, Ward~B Manchester, Alfred~O Hero, Gabor Toth, Benoit DuFumier, Tian
  Zhou, Xiantong Wang, Haonan Zhu, Zeyu Sun, and Tamas~I Gombosi.
\newblock Identifying solar flare precursors using time series of sdo/hmi
  images and sharp parameters.
\newblock \emph{Space Weather}, 17\penalty0 (10):\penalty0 1404--1426, 2019.

\bibitem[Cohen(1960)]{cohen1960coefficient}
Jacob Cohen.
\newblock A coefficient of agreement for nominal scales.
\newblock \emph{Educational and psychological measurement}, 20\penalty0
  (1):\penalty0 37--46, 1960.

\bibitem[Dikpati and Gilman(2006)]{dikpati2006simulating}
Mausumi Dikpati and Peter~A Gilman.
\newblock Simulating and predicting solar cycles using a flux-transport dynamo.
\newblock \emph{The Astrophysical Journal}, 649\penalty0 (1):\penalty0 498,
  2006.

\bibitem[Florios et~al.(2018)Florios, Kontogiannis, Park, Guerra, Benvenuto,
  Bloomfield, and Georgoulis]{florios2018forecasting}
Kostas Florios, Ioannis Kontogiannis, Sung-Hong Park, Jordan~A Guerra, Federico
  Benvenuto, D~Shaun Bloomfield, and Manolis~K Georgoulis.
\newblock Forecasting solar flares using magnetogram-based predictors and
  machine learning.
\newblock \emph{Solar Physics}, 293\penalty0 (2):\penalty0 1--42, 2018.

\bibitem[Garcia(1994)]{garcia1994temperature}
Howard~A Garcia.
\newblock Temperature and emission measure from goes soft x-ray measurements.
\newblock \emph{Solar Physics}, 154\penalty0 (2):\penalty0 275--308, 1994.

\bibitem[{Guerra} et~al.(2020){Guerra}, {Murray}, {Shaun Bloomfield}, and
  {Gallagher}]{Guerraetal2020}
Jordan~A. {Guerra}, Sophie~A. {Murray}, D.~{Shaun Bloomfield}, and Peter~T.
  {Gallagher}.
\newblock Ensemble forecasting of major solar flares: methods for combining
  models.
\newblock \emph{Journal of Space Weather and Space Climate}, 10:\penalty0 38,
  2020.
\newblock \doi{10.1051/swsc/2020042}.

\bibitem[Hale et~al.(1919)Hale, Ellerman, Nicholson, and Joy]{hale1919magnetic}
George~E Hale, Ferdinand Ellerman, Seth~Barnes Nicholson, and Alfred~Harrison
  Joy.
\newblock The magnetic polarity of sun-spots.
\newblock \emph{The Astrophysical Journal}, 49:\penalty0 153, 1919.

\bibitem[Hathaway(2015)]{hathaway2015solar}
David~H Hathaway.
\newblock The solar cycle.
\newblock \emph{Living reviews in solar physics}, 12\penalty0 (1):\penalty0
  1--87, 2015.

\bibitem[Hecht-Nielsen(1992)]{hecht1992theory}
Robert Hecht-Nielsen.
\newblock Theory of the backpropagation neural network.
\newblock In \emph{Neural networks for perception}, pages 65--93. Elsevier,
  1992.

\bibitem[Hiremath(2008)]{hiremath2008prediction}
KM~Hiremath.
\newblock Prediction of solar cycle 24 and beyond.
\newblock \emph{Astrophysics and Space Science}, 314\penalty0 (1):\penalty0
  45--49, 2008.

\bibitem[{Hoeksema} et~al.(2014){Hoeksema}, {Liu}, {Hayashi}, {Sun}, {Schou},
  {Couvidat}, {Norton}, {Bobra}, {Centeno}, {Leka}, {Barnes}, and
  {Turmon}]{Hoeksemaetal2014}
J.~T. {Hoeksema}, Y.~{Liu}, K.~{Hayashi}, X.~{Sun}, J.~{Schou}, S.~{Couvidat},
  A.~{Norton}, M.~{Bobra}, R.~{Centeno}, K.~D. {Leka}, G.~{Barnes}, and
  M.~{Turmon}.
\newblock The {Helioseismic} and {Magnetic} {Imager} ({HMI}) vector magnetic
  field pipeline: Overview and performance.
\newblock \emph{Solar Physics}, 289:\penalty0 3483--3530, 2014.
\newblock \doi{10.1007/s11207-014-0516-8}.

\bibitem[Hoeksema(1995)]{hoeksema1995large}
J~Todd Hoeksema.
\newblock The large-scale structure of the heliospheric current sheet during
  the ulysses epoch.
\newblock \emph{The High Latitude Heliosphere}, pages 137--148, 1995.

\bibitem[Huang et~al.(2018)Huang, Wang, Xu, Liu, Li, and Dai]{huang2018deep}
Xin Huang, Huaning Wang, Long Xu, Jinfu Liu, Rong Li, and Xinghua Dai.
\newblock Deep learning based solar flare forecasting model. i. results for
  line-of-sight magnetograms.
\newblock \emph{The Astrophysical Journal}, 856\penalty0 (1):\penalty0 7, 2018.

\bibitem[Jiao et~al.(2020)Jiao, Sun, Wang, Manchester, Gombosi, Hero, and
  Chen]{jiao2020solar}
Zhenbang Jiao, Hu~Sun, Xiantong Wang, Ward Manchester, Tamas Gombosi, Alfred
  Hero, and Yang Chen.
\newblock Solar flare intensity prediction with machine learning models.
\newblock \emph{Space Weather}, 18\penalty0 (7):\penalty0 e2020SW002440, 2020.

\bibitem[Jonas et~al.(2018)Jonas, Bobra, Shankar, Todd~Hoeksema, and
  Recht]{jonas2018flare}
Eric Jonas, Monica Bobra, Vaishaal Shankar, J~Todd~Hoeksema, and Benjamin
  Recht.
\newblock Flare prediction using photospheric and coronal image data.
\newblock \emph{Solar Physics}, 293\penalty0 (3):\penalty0 1--22, 2018.

\bibitem[Kors{\'o}s et~al.(2021)Kors{\'o}s, Erd{\'e}lyi, Liu, and
  Morgan]{korsos2021testing}
MB~Kors{\'o}s, R~Erd{\'e}lyi, Jiajia Liu, and H~Morgan.
\newblock Testing and validating two morphological flare predictors by logistic
  regression machine learning.
\newblock \emph{Frontiers in Astronomy and Space Sciences}, 7:\penalty0 571186,
  2021.

\bibitem[{Leka} et~al.(2018){Leka}, {Barnes}, and {Wagner}]{Lekaetal2018}
K.~D. {Leka}, G.~{Barnes}, and E.~L. {Wagner}.
\newblock The nwra classification infrastructure: description and extension to
  the discriminant analysis flare forecasting system (daffs).
\newblock \emph{J.~Space Weather Space Clim.}, 8:\penalty0 A25, 2018.
\newblock \doi{10.1051/swsc/2018004}.

\bibitem[Leka et~al.(2019)Leka, Park, Kusano, Andries, Barnes, Bingham,
  Bloomfield, McCloskey, Delouille, Falconer, et~al.]{leka2019comparison}
KD~Leka, Sung-Hong Park, Kanya Kusano, Jesse Andries, Graham Barnes, Suzy
  Bingham, D~Shaun Bloomfield, Aoife~E McCloskey, Veronique Delouille, David
  Falconer, et~al.
\newblock A comparison of flare forecasting methods. ii. benchmarks, metrics,
  and performance results for operational solar flare forecasting systems.
\newblock \emph{The Astrophysical Journal Supplement Series}, 243\penalty0
  (2):\penalty0 36, 2019.

\bibitem[Liu et~al.(2017)Liu, Deng, Wang, and Wang]{liu2017predicting}
Chang Liu, Na~Deng, Jason~TL Wang, and Haimin Wang.
\newblock Predicting solar flares using sdo/hmi vector magnetic data products
  and the random forest algorithm.
\newblock \emph{The Astrophysical Journal}, 843\penalty0 (2):\penalty0 104,
  2017.

\bibitem[Liu and Nocedal(1989)]{liu1989limited}
Dong~C Liu and Jorge Nocedal.
\newblock On the limited memory bfgs method for large scale optimization.
\newblock \emph{Mathematical programming}, 45\penalty0 (1):\penalty0 503--528,
  1989.

\bibitem[Liu et~al.(2019)Liu, Liu, Wang, and Wang]{liu2019predicting}
Hao Liu, Chang Liu, Jason~TL Wang, and Haimin Wang.
\newblock Predicting solar flares using a long short-term memory network.
\newblock \emph{The Astrophysical Journal}, 877\penalty0 (2):\penalty0 121,
  2019.

\bibitem[Morgan and Taroyan(2017)]{morgan2017global}
Huw Morgan and Youra Taroyan.
\newblock Global conditions in the solar corona from 2010 to 2017.
\newblock \emph{Science advances}, 3\penalty0 (7):\penalty0 e1602056, 2017.

\bibitem[Murphy(2012)]{murphy2012machine}
Kevin~P Murphy.
\newblock \emph{Machine learning: a probabilistic perspective}.
\newblock MIT press, 2012.

\bibitem[Nishizuka et~al.(2017)Nishizuka, Sugiura, Kubo, Den, Watari, and
  Ishii]{nishizuka2017solar}
N~Nishizuka, K~Sugiura, Y~Kubo, M~Den, S~Watari, and M~Ishii.
\newblock Solar flare prediction model with three machine-learning algorithms
  using ultraviolet brightening and vector magnetograms.
\newblock \emph{The Astrophysical Journal}, 835\penalty0 (2):\penalty0 156,
  2017.

\bibitem[{Pesnell} et~al.(2012){Pesnell}, {Thompson}, and
  {Chamberlin}]{PesnellThompsonChamberlin2012}
W.~D. {Pesnell}, B.~J. {Thompson}, and P.~C. {Chamberlin}.
\newblock The solar dynamics observatory (sdo).
\newblock \emph{Solar Physics}, 275:\penalty0 3--15, 2012.
\newblock \doi{10.1007/s11207-011-9841-3}.

\bibitem[{Pesnell}(2012)]{Pesnell:2012}
W.~Dean {Pesnell}.
\newblock {Solar Cycle Predictions (Invited Review)}.
\newblock \emph{Solar Physics}, 281\penalty0 (1):\penalty0 507--532, November
  2012.
\newblock \doi{10.1007/s11207-012-9997-5}.

\bibitem[{Pesnell}(2020)]{Pesnell:2020}
W.~Dean {Pesnell}.
\newblock {Lessons learned from predictions of Solar Cycle 24}.
\newblock \emph{Journal of Space Weather and Space Climate}, 10:\penalty0 60,
  October 2020.
\newblock \doi{10.1051/swsc/2020060}.

\bibitem[{Scherrer} et~al.(2012){Scherrer}, {Schou}, {Bush}, {Kosovichev},
  {Bogart}, {Hoeksema}, {Liu}, {Duvall}, {Zhao}, {Title}, {Schrijver},
  {Tarbell}, and {Tomczyk}]{Scherreretal2012}
P.~H. {Scherrer}, J.~{Schou}, R.~I. {Bush}, A.~G. {Kosovichev}, R.~S. {Bogart},
  J.~T. {Hoeksema}, Y.~{Liu}, T.~L. {Duvall}, J.~{Zhao}, A.~M. {Title}, C.~J.
  {Schrijver}, T.~D. {Tarbell}, and S.~{Tomczyk}.
\newblock The {Helioseismic} and {Magnetic} {Imager} ({HMI}) investigation for
  the {Solar} {Dynamics} {Observatory} ({SDO}).
\newblock \emph{Solar Physics}, 275:\penalty0 207--227, 2012.
\newblock \doi{10.1007/s11207-011-9834-2}.

\bibitem[{Sun} et~al.(2012){Sun}, {Hoeksema}, {Liu}, {Wiegelmann}, {Hayashi},
  {Chen}, and {Thalmann}]{SunX:2012}
Xudong {Sun}, J.~Todd {Hoeksema}, Yang {Liu}, Thomas {Wiegelmann}, Keiji
  {Hayashi}, Qingrong {Chen}, and Julia {Thalmann}.
\newblock {Evolution of Magnetic Field and Energy in a Major Eruptive Active
  Region Based on SDO/HMI Observation}.
\newblock \emph{The Astrophysical Journal}, 748\penalty0 (2):\penalty0 77,
  April 2012.
\newblock \doi{10.1088/0004-637X/748/2/77}.

\bibitem[Sun et~al.(2022)Sun, Bobra, Wang, Wang, Sun, Gombosi, Chen, and
  Hero]{sun2022predicting}
Zeyu Sun, Monica~G Bobra, Xiantong Wang, Yu~Wang, Hu~Sun, Tamas Gombosi, Yang
  Chen, and Alfred Hero.
\newblock Predicting solar flares using cnn and lstm on two solar cycles of
  active region data.
\newblock \emph{The Astrophysical Journal}, 931\penalty0 (2):\penalty0 163,
  2022.

\bibitem[Svalgaard et~al.(1978)Svalgaard, Duvall, and
  Scherrer]{svalgaard1978strength}
Leif Svalgaard, Thomas~L Duvall, and Philip~H Scherrer.
\newblock The strength of the sun's polar fields.
\newblock \emph{Solar Physics}, 58\penalty0 (2):\penalty0 225--239, 1978.

\bibitem[Svalgaard et~al.(2005)Svalgaard, Cliver, and
  Kamide]{svalgaard2005sunspot}
Leif Svalgaard, Edward~W Cliver, and Yohsuke Kamide.
\newblock Sunspot cycle 24: Smallest cycle in 100 years?
\newblock \emph{Geophysical Research Letters}, 32\penalty0 (1), 2005.

\bibitem[Upton and Hathaway(2013)]{upton2013predicting}
Lisa Upton and David~H Hathaway.
\newblock Predicting the sun's polar magnetic fields with a surface flux
  transport model.
\newblock \emph{The Astrophysical Journal}, 780\penalty0 (1):\penalty0 5, 2013.

\bibitem[Wang et~al.(2020)Wang, Chen, Toth, Manchester, Gombosi, Hero, Jiao,
  Sun, Jin, and Liu]{wang2020predicting}
Xiantong Wang, Yang Chen, Gabor Toth, Ward~B Manchester, Tamas~I Gombosi,
  Alfred~O Hero, Zhenbang Jiao, Hu~Sun, Meng Jin, and Yang Liu.
\newblock Predicting solar flares with machine learning: Investigating solar
  cycle dependence.
\newblock \emph{The Astrophysical Journal}, 895\penalty0 (1):\penalty0 3, 2020.

\bibitem[Wang et~al.(2005)Wang, Lean, and Sheeley~Jr]{wang2005modeling}
Y-M Wang, JL~Lean, and NR~Sheeley~Jr.
\newblock Modeling the sun’s magnetic field and irradiance since 1713.
\newblock \emph{The Astrophysical Journal}, 625\penalty0 (1):\penalty0 522,
  2005.

\bibitem[Yuksel et~al.(2012)Yuksel, Wilson, and Gader]{yuksel2012twenty}
Seniha~Esen Yuksel, Joseph~N Wilson, and Paul~D Gader.
\newblock Twenty years of mixture of experts.
\newblock \emph{IEEE transactions on neural networks and learning systems},
  23\penalty0 (8):\penalty0 1177--1193, 2012.

\end{thebibliography}

\section*{Appendix A: Definitions of the SHARP parameters}

\begin{table}[h]
    \centering
    \small
    \ma{
    \begin{tabular}{m{2cm} |m{10cm} }
        \hline
        Parameters & Description  \\
        \hline
        USFLUXL & Total unsigned flux \\
        MEANGBL & Mean gradient of the line-of-sight field \\
        RVALUE & Total line-of-sight unsigned flux \\
        AREA & De-projected area of patch on sphere in microhemisphere \\
        TOTUSJH & Total unsigned current helicity \\
        TOTUSJZ & Total unsigned vertical current \\
        SAVNCPP & Sum of the modulus of the net current per polarity \\
        ABSNJZH & Absolute value of the net current helicity \\
        TOTPOT & Proxy for total photospheric magnetic free energy density \\
        SIZEACR & Deprojected area of active pixels (Bz magnitude larger than noise threshold) on image in microhemisphere (defined as one millionth of half the surface of the Sun) \\
        NACR & The number of strong LOS magnetic field pixels in the patch \\
        MEANPOT & Proxy for mean photospheric excess magnetic energy density \\
        SIZE & Projected area of the image in microhemispheres \\
        SHRGT45 & Fraction of area with shear $>$ 45\textdegree \\
        MEANSHR & Mean shear angle \\
        MEANJZD & Vertical current density \\
        MEANALP & Characteristic twist parameter, $\alpha$ \\
        MEANGBT & Horizontal gradient of total field \\
        MEANGAM & Mean angle of field from radial \\
        MEANGBH & Horizontal gradient of horizontal field \\
        NPIX &  The number of pixels in a SHARP image \\
        \hline
    \end{tabular}
    }
    \caption{List of SHARP Parameters and Brief Descriptions}
    \label{tab:datasetsapp}
\end{table}

\end{document}